\documentclass[pra,onecolumn,floatfix,superscriptaddress,longbibliography,notitlepage, nofootinbib]{revtex4-1}
\pdfoutput=1

\usepackage[utf8]{inputenc}
\usepackage{braket}
\usepackage{amsmath}
\DeclareMathOperator{\Tr}{Tr}
\usepackage{dcolumn}   
\usepackage{bm}        
\usepackage{amssymb}
\usepackage{mathtools}
\usepackage{tikz}
\usepackage{qcircuit}
\usepackage{appendix}
\usepackage{hyperref}

\usepackage{subcaption}
\usepackage{amsmath,amsfonts,amssymb}
\usepackage{mathtools}
\usepackage{thmtools,thm-restate}
\usepackage{algorithm}
\usepackage{mathrsfs}

\usepackage{algpseudocode}

\newtheorem{definition}{Definition}
\newtheorem{theorem}{Theorem}
\newtheorem{lemma}{Lemma}

\newcommand{\revise}[1]{\textcolor{black}{#1}}
\newcommand{\xbf}{\textbf{x}}

\newcommand{\Ibb}{\mathbb{I}}

\usetikzlibrary{arrows.meta}

\def\BibTeX{{\rm B\kern-.05em{\sc i\kern-.025em b}\kern-.08em
    T\kern-.1667em\lower.7ex\hbox{E}\kern-.125emX}}

\begin{document}
\title{New Quantum Algorithm for Principal Component Analysis }
\author{Nhat A. Nghiem}
\affiliation{Department of Physics and Astronomy, State University of New York at Stony Brook, Stony Brook, NY 11794-3800, USA}
\affiliation{C. N. Yang Institute for Theoretical Physics, State University of New York at Stony Brook, Stony Brook, NY 11794-3840, USA}

\begin{abstract}
    Quantum principal component analysis (QPCA) ignited a new development toward quantum machine learning algorithms. Initially showcasing as an active way for analyzing a quantum system using the quantum state itself, QPCA also found potential application in analyzing a large-scale dataset, represented by the so-called covariance matrix. Inspired by recent advancement in quantum algorithms, we give an alternatively new quantum framework for performing principal component analysis. By analyzing the performance in detail, we shall identify the regime in which our proposal performs better than the original QPCA. In addition, we also provide a new approach for preparing the covariance matrix, given classical dataset, on a quantum computer. Thus, our work provides an efficient complementary framework for revealing features of the quantum state, while keeping the philosophy of original QPCA, as the state could play an active role in analyzing itself. 
\end{abstract}
\maketitle

\section{Introduction}
Quantum principal component analysis was initially proposed in \cite{lloyd2013quantum} as a technique for performing density matrix simulation. It immediately stood out as one of the very early influential quantum machine learning algorithms, as it could be applied to analyze a large-scale dataset, given the ability to prepare the so-called covariance matrix encoded in a density state. For example, as discussed in \cite{lloyd2013quantum}, combining density matrix simulation technique and quantum phase estimation algorithm \cite{kitaev1995quantum} reveals the largest eigenvalue and the corresponding eigenvectors, also dubbed principal components, of the given state. Thus, it enabled what was referred to in \cite{lloyd2013quantum} as self-tomography.  While the suggestion in \cite{lloyd2013quantum} regarding how to prepare a desired covariance matrix, given classical data, seems simple, its actual realization is beyond near-term implementation, as it requires quantum random access memory \cite{giovannetti2008architectures,giovannetti2008quantum}. A more friendly protocol has been constructed in \cite{gordon2022covariance}, and even more strikingly, it was successfully shown in the same work that both the real-world dataset and the physical dataset, e.g., MNIST and the molecular ground-state dataset, could be analyzed using the (quantum) PCA technique. 

At the same time, the development of quantum algorithmic frameworks has seen major progress. Some early constructions, such as quantum search algorithm \cite{grover1996fast}, quantum factorization algorithm \cite{shor1999polynomial}, computing black-box properties \cite{deutsch1985quantum,deutsch1992rapid}, etc., have greatly influenced the search for quantum computational advantage. Subsequently, many seminal quantum algorithms have been constructed to tackle a wide array of computational challenges. Within the context of physics, as fueled by an early proposal \cite{feynman2018simulating,lloyd1996universal}, a series of attempts \cite{berry2007efficient,berry2012black,berry2014high,berry2015hamiltonian, childs2010relationship, low2017optimal,low2019hamiltonian, berry2020time, watkins2024time, chen2021quantum, an2022time, an2021time, kieferova2019simulating, low2018hamiltonian, ambainis2007quantum} have been made to simulate a time-independent and time-dependent Hamiltonian. In particular, the success of the quantum simulation algorithm has also impacted further quantum algorithm applications. For instance, building the Hamiltonian simulation technique, \cite{harrow2009quantum, childs2017quantum} have constructed a quantum linear solver with exponential speedup compared to the classical solver. Additionally, the authors in \cite{wiebe2012quantum} have shown that a simple adaption of the quantum linear solver could perform least-square data fitting, which is a major problem of great importance in machine learning and data science. In particular, the success of the quantum linear solving algorithm also underlies many more quantum ordinary differential equation and partial differential equation solvers \cite{berry2014high, leyton2008quantum, liu2021efficient, berry2017quantum, tennie2024quantum}. Most recently, a class of quantum algorithms, the so-called variational quantum algorithm \cite{mitarai2018quantum,schuld2014quest,schuld2018supervised,schuld2019evaluating,schuld2019machine,schuld2019quantum,schuld2020circuit, havlivcek2019supervised, liu2021rigorous}, has been proposed as a promising candidate to run on near-term quantum computers, as it would circumvent the limitation of depth and hardware error. 

The aforementioned examples above have shown exciting progress of quantum algorithms, as a different purpose would lead to a different quantum algorithmic technique. However, a major cornerstone has recently been placed in the seminal work \cite{gilyen2019quantum}, so-called quantum singular value transformation (QSVT) framework. It turns out that many quantum algorithms could be sitting on a common ground, as QSVT allows a unified description of many prior techniques. Additionally, as was shown in \cite{gilyen2019quantum}, many quantum algorithms, e.g., quantum walk, admit a simplified execution, mostly utilizing single qubit rotation gates. Thus, QSVT changes the landscape of quantum algorithms in a sense that researchers/learners could approach the topic of quantum algorithms in a simpler, yet potentially more intuitive fashion. An intriguing question stemming from the construction of QSVT is that, aside from the representation itself, could QSVT enable any stronger means to deal with computational problems. Some works have affirmed it to be positive. For example, \cite{gilyen2022quantum} have shown how to use QSVT techniques to construct the Petz recovery channel, which is a pivotal tool in quantum information science, and that prior to QSVT framework, there was no systematic way to construct it. The work \cite{nghiem2023improved1} has suggested using simple tools from QSVT to construct a greatly improved version of quantum gradient descent, which was first introduced in \cite{rebentrost2019quantum}. In \cite{mitarai2023perturbation}, the authors also demonstrated that QSVT could be employed to estimate pertubative energies, based on pertubation theory.

The above success of demonstrating an enormous capability of quantum singular value transformation has indeed motivated us to go further. The particular problem that we encounter in this line of pursuit is principal component analysis (PCA). The quantum algorithm for performing the principal component analysis has been proposed in \cite{lloyd2013quantum}. It marked a very early and influentially major cornerstone of quantum machine learning algorithm, as PCA is one of the central tools within machine learning and big data science. Typically, in these areas, data points are usually of a high dimension and possess significant challenges for analyzing using classical resources. Meanwhile, the intrinsic nature of the quantum world allows information, in theory, to be stored using (poly)logarithmical resources. Indeed, as shown in \cite{lloyd2013quantum}, under certain assumptions, quantum computer could reveal the ``key'' features, also called principal components, of the covariance matrix that captures the correlation between different features of data. The technique underlying \cite{lloyd2013quantum} is the exponentiation of the density matrix and, in fact, it even has a major application within the physical context. More concretely, \cite{lloyd2013quantum} showed that quantum PCA could also be used to reveal the features of the quantum state themselves. Thus, it allows what was referred to as self-tomography, which is distinct from the standard tomography approaches relying on quantum state measurement. Our work revisits the (quantum) PCA problem with some inspiration from the seminal quantum singular value transformation framework. The hope is to see whether or not QSVT can bring new light to the originally proposed quantum PCA. As we shall see later, the answer turns out to be affirmatively yes. By analyzing the performance in detail, we would show that our new construction of the quantum principal component analysis algorithm could perform extremely well in the regime where the original quantum PCA \cite{lloyd2013quantum} performs the worst. In addition, we also revisit a related work \cite{gordon2022covariance} using our new perspective and our new quantum PCA algorithm. The work \cite{gordon2022covariance} initially aimed to fill the gap from \cite{lloyd2013quantum}, as the protocol for preparing the covariance matrix in \cite{lloyd2013quantum} was not explicit. We will show that the preparation of the covariance matrix could be achieved in a simple manner using tools borrowed from QSVT \cite{gilyen2019quantum}. In particular, as a byproduct, our protocol for covariance matrix preparation can naturally deal with uncentered dataset (to be defined later), thus it proves to be slightly more useful than the protocol of \cite{gordon2022covariance}, as we would not lose the contribution from the so-called centroids, for which we will elaborate further in the same context. 

The structure of our work is as follows. In section \ref{sec: review}, we give a revision of the original quantum PCA algorithm \cite{lloyd2013quantum}, with key techniques and corresponding complexity explicitly stated. Then subsequently in sections \ref{sec: quantumalgorithm} and \ref{sec: fewlargest}, we outline our main result, which is a new quantum algorithm to perform the analysis of principal components. We remark that a lot of recipes we would use are borrowed from QSVT context \cite{gilyen2019quantum}, thus we refer to the original work for more details, however, in appendix \ref{sec: prelim}, we recapitulate precisely all techniques used in our work. In section \ref{sec: comparison}, we provide a comparison between our proposal and the original quantum PCA. We explicitly discuss the corresponding scenario of the best performance of two quantum algorithms, whereby we conclude that the two frameworks complement each other very well. Section \ref{sec: application} is devoted to the preparation of a covariance matrix, where we first provide an overview of PCA setting as well as related concepts, including centroids, the covariance matrix and, in particular, a scenario of ``PCA without centering''. We then outline two alternative ways to prepare the desired covariance matrix, given access to classical data. A brief summary is provided in Section \ref{sec: outlook}, following with comments on the prospect of our framework from a broader context, including practical implication and potential quantum advantage.

\section{Review of Previous Quantum Principal Component Analysis Algorithm}
\label{sec: review} 
The original procedure provided in \cite{lloyd2013quantum} is as follows. Suppose that we are given multiple copies of the quantum state $\rho \in \mathbb{C}^{n \times n}$ where $n$ is the dimension. In order to obtain the unitary transformation $\exp(- i \rho t)$, the authors in \cite{lloyd2013quantum} used the following property:
\begin{align}
    \Tr_1 \exp(-iS \Delta t) \big( \rho \otimes \sigma \big) \exp(-iS \Delta t) = \sigma - i \Delta t [\rho,\sigma] + \mathcal{O}(\Delta t^2) \approx \exp(-i \rho \Delta t) \sigma \exp(-i \rho \Delta t)
\end{align}
where $\Tr_1$ is the partial trace over the first system, $S$ is the swap operator between two system of $\log(n)$ qubits, and $\sigma$ is some ancilla system. Following the above step:
\begin{align}
     \Tr_1 \exp(-iS \Delta t)  \big( \rho \otimes ( \exp(-i \rho \Delta t) \sigma \exp(-i \rho \Delta t))   \big) \approx \exp(-\rho \Delta t) ( \exp(-i \rho \Delta t) \sigma \exp(-i \rho \Delta t) ) \exp(-i \rho \Delta t)
\end{align}
In order to obtain the transformation $\exp(-i \rho t)$, the above procedure is repeated $N$ times, and we have approximated obtained:
\begin{align}
    \exp(-i \rho N \Delta t) \sigma \exp(-i \rho N \Delta t)
\end{align}
The authors in \cite{lloyd2013quantum} shows that to simulate $\exp(-i \rho t)$  to accuracy $\epsilon$, then it requires:
\begin{align}
    N = \mathcal{O} \big(  \frac{t^2}{\epsilon}  \big)
\end{align}
where $t = N\Delta t$. We summarize the above procedure in the following lemma:
\begin{lemma}[Density matrix exponentiation]
\label{lemma: dme}
    There exists a procedure composed of a quantum circuit of depth $\mathcal{O}( \frac{t^2}{\epsilon}\log(n))$ and $N = \mathcal{O}(t^2/\epsilon)$ copies of given state $\rho$ of dimension $n\times n$, that approximate the unitary $\exp(-i\rho t)$ to accuracy $\epsilon$. 
\end{lemma}
The above procedure yields (approximately) the operator $\exp(-i \rho t)$. In order to perform the principal component analysis, we need to combine it with the quantum phase estimation algorithm \cite{kitaev1995quantum, brassard2002quantum}. If we perform phase estimation algorithm with unitary $\exp(i \rho t)$ as the main unitary that we wish to reveal the phase, and $\rho$ as input state, then we obtain the following:
\begin{align}
\label{5}
    \sum_i r_i \ket{\lambda_i}\bra{\lambda_i} \otimes \ket{\Tilde{r}_i} \bra{\Tilde{r}_i}
\end{align}
where $\{ \lambda_i \}_i$ are eigenvectors of $\rho$ and $\{ r_i, \Tilde{r}_i \}_i$ are corresponding eigenvalues/approximated eigenvalues of $\rho$. The reason we have $\Tilde{r}_i$ is that quantum phase estimation yields an approximation of eigenvalues. According to \cite{lloyd2013quantum}, if we wish that the error resulting from the phase estimation algorithm is $\mathcal{O}(\epsilon)$ then we need to choose $t = \mathcal{O}(1/\epsilon)$ in the quantum phase estimation algorithm. The total number of copies required is $N = \mathcal{O}(1/\epsilon^3)$ obtain the above state. In order to reveal the features, for example, if we wish to obtain the eigenvector of $\rho$ with the largest eigenvalue, we need to perform a measurement in the phase register of the above state. 

We discuss a few aspects of the quantum principal component algorithm above. If we wish to obtain the eigenstate that corresponds to the largest eigenvalue, the success probability is then $|r_{\max}|^2$ where $r_{\max} = \max_i \{ r_i \}$, and hence the algorithm is the most efficient when $r_{\max}$ dominate the remaining eigenvalues. The overall complexity, including the number of copies of $\rho$ used and a quantum circuit depth required, is respectively:
\begin{align*}
    \mathcal{O}\big(  \frac{1}{r_{\max}^2 \epsilon^3} \big) , \mathcal{O}\big(  \frac{1}{r_{\max}^2 \epsilon^3} \log(n) \big)
\end{align*}
In addition, if we wish to estimate the largest eigenvalue $r_{\max}$, then we need to repeat the procedure further $\mathcal{O}(1/ \epsilon^2)$ times to obtain the estimation up to additive error $\epsilon$, resulting in a total running time $\mathcal{O}(1/ \epsilon^5)$. If, aside from the largest eigenvalue/eigenvector, we also care about, say, $k$ largest eigenvalue/eigenvector, then the only solution is to keep measuring the phase register and select the largest $k$ eigenvalues, with corresponding eigenvectors. As also emphasized in \cite{lloyd2013quantum}, quantum principal component analysis is most efficient when $\rho$ is dominated by $k$ largest few eigenvalues. As an example, the authors in \cite{lloyd2013quantum} suggested the scenario where the density matrix $\rho$ is corresponding with covariance matrix of some dataset.

\section{New Quantum Algorithm for Principal Component Analysis}
\label{sec: quantumalgorithm} 
Our method begins with a simple recipe from the quantum singular value transformation framework \cite{gilyen2019quantum}:
\begin{lemma}[Logarithmic of Unitary, Corollay 71 in \cite{gilyen2019quantum}]
\label{lemma: logarithmicofunitary}
    Suppose that $U = \exp(-iH)$, where $H$ is a Hamiltonian of norm at most $1/2$. Let $\epsilon \in (0,1/2]$, then we can implement an $\epsilon$-approximated block encoding of $2H/\pi$ (see further definition \ref{def: blockencode}) with $\mathcal{O}(\log(\frac{1}{\epsilon} ))$ uses of controlled-U and its inverse, using $\mathcal{O}(\log(\frac{1}{\epsilon}))$ two-qubit gates and using a single ancilla qubit. 
\end{lemma}
Given that in previous section we have obtained the unitary $\exp(-i \rho t)$ up to additive accuracy $\epsilon$, then we can apply the above result in a straightforward manner. However, there are two issues that we need to take into account. First, in the above lemma, $H$ is required to have the norm less than $1/2$, meanwhile, in our case, $\rho$ is generally having norm less than $1$. Second, in the context of the above lemma, the unitary $U$ is ideal, meanwhile, in our case, $U$ is constructed with additive error $\epsilon$. To resolve the first issue, we simply need to choose $t =1/2$, i.e., we construct the block encoding of $\exp(-i \rho/2)$. For the second issue, we need to go deeper into the mechanism underlying the above lemma, which is also the central technique from QSVT: 
\begin{lemma}\label{lemma: qsvt}[\cite{gilyen2019quantum} Theorem 56]
\label{lemma: theorem56} 
Suppose that $U$ is an
$(\alpha, a, \epsilon)$-encoding of a Hermitian matrix $A$. (See Definition 43 of~\cite{gilyen2019quantum}, or definition \ref{def: blockencode} for the definition.)
If $P \in \mathbb{R}[x]$ is a degree-$d$ polynomial satisfying that
\begin{itemize}
\item for all $x \in [-1,1]$: $|P(x)| \leq \frac{1}{2}$,
\end{itemize}
then, there is a quantum circuit $\tilde{U}$, which is an $(1,a+2,4d \sqrt{\frac{\epsilon}{\alpha}})$-encoding of $P(A/\alpha)$ and
consists of $d$ applications of $U$ and $U^\dagger$ gates, a single application of controlled-$U$ and $\mathcal{O}((a+1)d)$
other one- and two-qubit gates.
\end{lemma}
In order to apply the above lemma to the context of lemma \ref{lemma: logarithmicofunitary}, as shown in \cite{gilyen2019quantum}, choosing $P$ to be a degree-$d=\mathcal{O}(\log\frac{1}{\delta})$ polynomial is sufficient to approximate the desired polynomial of accuracy $\delta$, and we fix this $\delta$ to be some constant. In our context, we point out that, as any unitary block encodes itself, we can apply the above lemma directly to such a unitary. As the unitary $U = \exp(-i \rho/2)$ is realized with additive error $\epsilon$ (from lemma \ref{lemma: dme}), then we would obtain an approximation to the block encoding of $\pi(\rho/2) /2  = \pi\rho/4$. The error approximation, according to the lemma \ref{lemma: theorem56} is $4d\sqrt{\epsilon} = \mathcal{O}( \sqrt{\epsilon}) $ where we have set $d = \mathcal{O}(1/\delta)$ to be some constant (independent of system size). For simplicity, we set the overall error to be $\epsilon$ then we need to rescale the error as:
\begin{align*}
    \sqrt{\epsilon}  \longrightarrow \epsilon
\end{align*}
which means that we need to use lemma \ref{lemma: dme} with accuracy $\epsilon^2$, and hence the number of copies of $\rho$ is $N = \mathcal{O}(1/\epsilon^2)$ would be sufficient to construct an $\epsilon$-approximated block encoding of $\pi \rho /4$. We summarize the above tool that we need for subsequent construction in the following lemma:
\begin{lemma}
    \label{lemma: unitaryblockencodingrho}
    There is a procedure using $\mathcal{O}( \frac{1}{\epsilon^2} )$ copies of $\rho$ and a quantum circuit of depth $\mathcal{O}( \log(n) \frac{1}{\epsilon^2} )$ that output the $\mathcal{O}( \epsilon)$-approximated block encoding of $\frac{\pi \rho}{4}$ 
\end{lemma}

Now we proceed with the next part of our algorithm, which concerns the largest eigenvalues and eigenvectors of $\rho$, or more precisely, of $\frac{\pi \rho}{4}$. The idea is built upon the (classical) power method, for which a quantum version (within oracle model) has been proposed in \cite{nghiem2023quantum}. Power method has a very simple execution, which is constantly applying the given matrix to an initially random state, and estimate the largest eigenvalue from a simple overlap estimation. The algorithm is formalized as following.
\begin{lemma}[Classical Power Method]
\label{lemma: classicalpowermethod}
    Given a Hermitian matrix $A \in \mathbb{C}^{n\times n}$ and a random vector $x_0 \in \mathbb{C}^n$. Define 
    \begin{align*}
        x_k = A^k x_0, \text{ and } \ket{x_k} = \frac{x_k}{|x_k|_2}
    \end{align*}
    where $|.|_2$ refers to $l_2$ Euclidean norm. Then for $k = \mathcal{O}( \log \frac{1}{\epsilon})  $. Let $\{r_i\}_{i=1}^n$ be the eigenvalue of $A$ with order $|r_1| > |r_2| > ... > |r_n|$ and $\{\lambda_i \}_{i=1}^n$ be the corresponding eigenvectors of $A$ (assumed to have unit norm for simplicity). Let $\gamma$ be the gap (absolute difference in magnitude) between the two largest eigenvalues of $A$, $| r_1 - r_2 |$, then we have for $k = \mathcal{O}( \frac{1}{\gamma} \log \frac{1}{\delta})$: \begin{align*} | \bra{x_k} A \ket{x_k} - r_1 | \leq \delta
    \end{align*}
\end{lemma}
We remark that, as analyzed in \cite{nghiem2023quantum, nghiem2023improved1, nghiem2024improved}, the original proposal of quantum power method in \cite{nghiem2023quantum} suffers from measurement issue (the probability of obtaining a desired state grows exponentially small), making the quantum power method only gain quadratic speedup in the number of dimension $n$. In the same works, \cite{nghiem2023improved1, nghiem2024improved} provided improved quantum power methods upon that of \cite{nghiem2023quantum} that recovers exponential speedup. On one hand, the method in \cite{nghiem2023improved1} uses a trick to reduce the eigenvalue finding to solving a small linear system size, thus avoiding the measurement step, which possesses an exponential barrier. On the other hand, the method introduced by the same authors in \cite{nghiem2024improved} utilizes the recently introduced quantum singular value transformation framework, thus providing a more direct way to improve the original quantum power method. 

The key difference from two works \cite{nghiem2023improved1,nghiem2024improved} and our work is that we are not assuming an oracle model (e.g., the oracle access to entries of matrix $A$ in the above lemma). However, the central idea of these improvements \cite{nghiem2023improved1, nghiem2024improved} can still be applied directly in our work. To proceed, we first summarize the main result of \cite{nghiem2024improved, nghiem2023improved1} in the following lemma:
\begin{lemma}
\label{lemma: improvedpowermethod}
    Given a (or could be an $\epsilon$-approximated) block encoding (with $T_A$ as complexity, e.g., quantum circuit depth) of some Hermitian operator $A$ (as defined in previous lemma \ref{lemma: classicalpowermethod}), then there is a quantum procedure with quantum circuit depth $\mathcal{O}(  T_A \frac{1}{\gamma} \log\frac{1}{\epsilon}) $ that output a quantum state $\ket{\Tilde{x} }$ such that 
    $$ |\ket{\Tilde{x}    } - \lambda_1  | \leq \epsilon $$
    Additionally, by using the above procedure plus the block encoding of $A$ further $\mathcal{O}(1/\epsilon)$ times, then the largest eigenvalue of $A$, $r_1$ could be estimate with accuracy $\epsilon$.
\end{lemma}
We remark that as $\rho$ is a density matrix, it is both positive and Hermitian, which means that all eigenvalues having positive sign. Thus, the prior construction of the block encoding of $\pi\rho/4$ is a nice complement to the above result and vice versa. However, there is a subtle detail that we need to take into account, concerning the error. As we mentioned, our work doesn't belong oracle model as in \cite{nghiem2023improved1,nghiem2024improved}, while still being able to execute their central algorithm. One of the steps appeared in \cite{nghiem2023improved1, nghiem2024improved} is taking the approximated block encoding of operator $A$ (which is obtainable in logarithmical steps using oracle access to $A$), and use lemma \ref{lemma: product} to construct the block encoding of $A^k$ for an integer $k$. Suppose that each unitary block encoding is an $\delta$-approximated block encoding of $A$, then the block encoding of $A^k$ would have an error to be $k \delta$. In order to bound the error as $\epsilon$ for convention, we need to set $k\delta = \epsilon$, and hence induces more complexity for the encoding of $A$, but as such encoding step is logarithmical, the gain is modest. However, in our case, as we see from lemma \ref{lemma: dme}, it takes $\mathcal{O}(1/\epsilon^2)$ copies of $\rho$ (plus a quantum circuit of depth $\mathcal{O}(\log(n))$ to obtain an $\epsilon$-approximated block encoding of $\pi\rho/4$. Hence, if we execute the same procedure (underlying lemma \ref{lemma: improvedpowermethod}), there is also a step where we need to use lemma \ref{lemma: product} to construct the block encoding of $(\pi\rho/4)^k$, and the accumulated error would be $k \epsilon$. We need such error to be $\epsilon$ (for convention); then we need to rescale $k\epsilon \longrightarrow \epsilon$. Hence, it actually takes $\mathcal{O}( k^2/\epsilon^2 )$ copies of $\rho$ to first construct the $\epsilon/k$-approximated block encoding of $\pi\rho/4$, and then construct the $\epsilon$-approximated block encoding of $(\pi\rho/4)^k$. Then the rest obey what have been outlined in \cite{nghiem2024improved, nghiem2023improved1}, as summarized in the above lemma \ref{lemma: improvedpowermethod}. We formally summarize our density matrix's largest eigenvalue finding algorithm as follows.
\begin{theorem}[Largest Component of Density Matrix $\rho$]
\label{theorem: improvedqpca}
    Suppose we are given multiple copies of a density state $\rho \in \mathbb{C}^{n\times n}$, with spectrum defined similarly to lemma \ref{lemma: classicalpowermethod}. Define $\gamma = |r_1-r_2|$ being the difference between two largest eigenvalue. There exists a quantum algorithm using a quantum circuit of depth $\mathcal{O}( \log(n) \frac{1}{\gamma^3} \log^3 (\frac{1}{\epsilon} )\frac{1}{\epsilon^2}  )$, using $N = \mathcal{O}( \frac{1}{\gamma^2} \log^2 (\frac{1}{\epsilon}) \frac{1}{\epsilon^2}   )$ copies of $\rho$ that output a estimation $\Tilde{r}$ and a quantum state $\ket{\Tilde{x}}$ such that:
    \begin{align*}
        |\Tilde{r} - r_1 | \leq \epsilon \\ 
        | \ket{\Tilde{x}} - \ket{\lambda_1} |_2 \leq \epsilon
    \end{align*}
\end{theorem}

\section{Discussion and Some Potential Extension}
\label{sec: discussion}
This section extends the discussion we had in the previous section to a broader context. Specifically, in the first part Sec.\ref{sec: fewlargest}, we begin by showing how to simply modify the algorithm in previous section to expand the reach to a few largest eigenvalues/eigenvectors. 

\subsection{From largest eigenvalue/eigenvector to a few largest ones}
\label{sec: fewlargest}
The above construction (theorem \ref{theorem: improvedqpca}) yields an approximation to the largest eigenvalue $r_1$ and the corresponding eigenvector $\ket{\lambda_1}$ of the density state. As also discussed in \cite{lloyd2013quantum}, one may also be interested in, for example, $R$ largest ones (the principal components). As $\rho$ is a positive matrix with unit trace with eigenvalues assumed to have descending order $r_1 > r_2 > ... > r_n$ (note that we are not restricting to low rank cases, and thus some $r_i$ can be zero), we have that $\rho - r_1 \ket{\lambda_1}\bra{\lambda_1}$ would have the eigenvalues to be $r_2 > r_3 > ... > r_n$. We recall that from the above discussion, (above theorem \ref{theorem: improvedqpca}), we need to rescale the error accordingly so as to apply power method with $k$ iterations, hence it takes $\mathcal{O}(k^2/\epsilon^2)$ copies of $\rho$ to construct the $\epsilon/k$-approximated block encoding of $\pi\rho/4$. Given that Theorem \ref{theorem: improvedqpca} yields the largest eigenstate $\ket{\lambda_1}$, we can apply the same procedure (as the procedure of Lemma \ref{lemma: unitaryblockencodingrho}) with $\ket{\lambda_1}\bra{\lambda_1}$ as input. Then $\mathcal{O}(k^2/\epsilon^2)$ copies of $\ket{\lambda_1}$ would be taken to obtain the $\epsilon/k$-approximated block encoding of $\pi \ket{\lambda_1}\bra{\lambda_1}/4$. Given such a block encoding and the (approximated) value of $r_1$, we can use the lemma \ref{lemma: scale} to construct the block encoding of $\pi r_1 \ket{\lambda_1}\bra{\lambda_1}/4$. Given that we already have a block encoding of $\pi\rho/4$, then we can use lemma \ref{lemma: sumencoding} to construct the $\epsilon/k$-approximated block encoding of:
\begin{align}
    \frac{\pi}{4}  \frac{1}{2}(\rho - r_1 \ket{\lambda_1}\bra{\lambda_1})
\end{align}
Then we can apply the same algorithm as theorem \ref{theorem: improvedqpca} to find the largest eigenvalue and corresponding eigenvectors of $\frac{\pi}{4} (\rho - r_1 \ket{\lambda_1}\bra{\lambda_1}) $. As mentioned, the largest eigenvalue of $\rho - r_1 \ket{\lambda_1}\bra{\lambda_1} $ is $r_2$. Hence, we can apply the algorithm of \ref{theorem: improvedqpca} to find $r_2$ and the corresponding eigenvector $\ket{\lambda_2}$. To find the next eigenvalue/eigenvector $r_3, \ket{\lambda_3}$, we again note that the matrix $\rho - r_1 \ket{\lambda_1}\bra{\lambda_1} - r_2 \ket{\lambda_2}\bra{\lambda_2}$ has its largest eigenvalue to be $r_3$. Hence, we can apply the same procedure in an iterative manner to proceed the finding, for example, if we wish to find $R$ largest eigenvalues/eigenvectors, corresponding with the principal components. We formalize the iterative procedure to find $R$ principal components in the following:
\begin{theorem}[Finding $R$ principal components]
\label{theorem: principalcomponents}
    Let $\rho,\gamma$ be defined as in \ref{theorem: improvedqpca} and that we want to find its principal components -- $R$ pairs largest eigenvalues/eigenvectors. There exists a quantum algorithm using 
    $$ N = \mathcal{O}\big(  (\frac{1}{\gamma^2} \log^2 (\frac{1}{\epsilon}) \frac{1}{\epsilon^2}  )^R \big)$$
    copies of $\rho$ and a quantum circuit of depth $\mathcal{O }\big( R \log(n) \frac{1}{\gamma^3} \log^3 (\frac{1}{\epsilon} )\frac{1}{\epsilon^2}   \big)$
\end{theorem}

\subsection{Comparison to original QPCA}
\label{sec: comparison}
This section is devoted for the purpose of discussing our improved quantum principal component algorithm in more detail, by providing a comparison between our work and the original work \cite{lloyd2013quantum}. As provided from section \ref{sec: review}, the number of copies required and the quantum circuit depth required for the orignal QPCA algorithm to output the largest eigenvalue and  corresponding eigenvector are, respectively:
\begin{align*}
    \mathcal{O}\big(  \frac{1}{r_{\max}^2 \epsilon^3} \big) , \mathcal{O}\big(  \frac{1}{r_{\max}^2 \epsilon^3} \log(n) \big)
\end{align*}
In terms of the number of copies required, our method (theorem \ref{theorem: improvedqpca}) achieves a slightly better dependence on inverse of error tolerance $\epsilon$. The seemingly two tricky factors to put into comparison are $\gamma$, which is the gap between two largest eigenvalues and $r_{\max} \equiv r_1$ (as we set from lemma \ref{lemma: classicalpowermethod} for convenience). Quite surprisingly, it turns out that our result complements very well that of \cite{lloyd2013quantum} in this regard. For example, as discussed in \cite{lloyd2013quantum}, their method would be more useful, probably most useful when $\rho$ could be represented accurately using its $R$ principal components, for example, the rank of $\rho$ is small. In such a case, only $R$ eigenvalues/eigenvectors are included in the decomposition \ref{5}, and the distribution of eigenvalues is uniform (each eigenvalue has a magnitude of roughly $1/R$), then we can sample these $R$ eigenstates/eigenvalues with equal probability. Hence, the largest eigenvalue/eigenstate could be yielded in overall complexity: 
\begin{align*}
    \mathcal{O}\big(  \frac{R }{\epsilon^3} \big) , \mathcal{O}\big(  \frac{R }{ \epsilon^3} \log(n) \big)
\end{align*}
On the other hand, our method has sample and depth complexity scaling as:
\begin{align*}
    \mathcal{O}( \frac{1}{\gamma^2} \log^2 (\frac{1}{\epsilon}) \frac{1}{\epsilon^2}   ), \mathcal{O}( \log(n) \frac{1}{\gamma^3} \log^2 (\frac{1}{\epsilon} )\frac{1}{\epsilon^3}  )
\end{align*}
where $\gamma$ is the gap between $r_1$ and $r_2$ -- two largest eigenvalues. Therefore, the regime where \cite{lloyd2013quantum} works best (uniform distribution of eigenvalues) is the regime where our method works less efficiently, as in such a case the gap $\gamma$ would be extremely small. Meanwhile, if there is a sufficiently large gap, e.g. $\mathcal{O}(1)$, between the largest eigenvalue and the second largest (also the remaining eigenvalues), then the method in \cite{lloyd2013quantum} would only works well in the case that the value of largest eigenvalue $r_1$ is sufficiently large. 

Now we turn attention to the case of outputting $R$ largest eigenvalues/eigenvectors, or the principal components, which should be a straightforward extension to the above discussion. As provided in theorem \ref{theorem: principalcomponents}, the number of copies of $\rho$ required has an exponential scaling with respect to $R$, therefore, our method only works well in the case where $R$ is very small, which is very possible in the case where $\rho$ has extremely low rank, for which $\rho$ could be effectively represented by these principal components. However, even when $\rho$ has a high rank, we can still choose to reveal the top $R$ components only for a very small $R$. As complexity depends only on the gap $\gamma$, the regime in which our method should perform the best is when $\gamma \in \mathcal{O}(1)$. As evident as we can see from the above complexity, high-rank regime impose significant difficulty for the original quantum PCA. 

Recall that the original QPCA \cite{lloyd2013quantum} has sample and quantum circuit depth complexity being 
\begin{align*}
    \mathcal{O}\big(  \frac{R }{\epsilon^3} \big) , \mathcal{O}\big(  \frac{R }{ \epsilon^3} \log(n) \big)
\end{align*}
in order to reveal $R$ largest eigenvalues plus their corresponding eigenvectors. If the gap between $R$ eigenvalues are big, then the method in \cite{lloyd2013quantum} could only output its $R$ eigenvalues/eigenvectors in complexity 
\begin{align*}
    \mathcal{O}\big(  \frac{r_{\min} }{\epsilon^3} \big) , \mathcal{O}\big(  \frac{r_{\min} }{ \epsilon^3} \log(n) \big)
\end{align*}
where $r_{\min} \equiv \min \{r_1,r_2,..., r_R\} $. Thus, for example, even if the density matrix $\rho$ has a considerably low rank and its minimum eigenvalue is exponentially small, then the original QPCA would still have very high complexity. As discussed above, this is exactly the regime in which our method potentially works best, as long as the gap $\gamma \in \mathcal{O}(1)$. The high rank , for example, $R \in \mathcal{O}(n)$, imposes a significant complexity on the original quantum PCA.

\section{New Method for Preparing Covariance Matrix}
\label{sec: application}
The original proposal of the analysis of principal quantum components initially focused on simulating the density matrix $\exp(-i \rho t)$ and using quantum phase estimation to perform the so-called self-tomography, revealing its own features. However, as discussed in \cite{lloyd2013quantum}, one of the central applications of QPCA is to analyze a large-scale dataset, described by a covariance matrix $A$. To be more specific, in a typical context of (classical) PCA, a set of $N$ data points $\{\xbf^i\}_{i=1}^N$, where each point $\xbf^i \in \mathbb{C}^n$ is a vector of dimensions $n$. The empirical mean of the data is the so-called centroid:
\begin{align}
    \mu = \frac{1}{N} \sum_{i=1}^N \xbf^i
\end{align}
which holds for the uniform distribution setting. Generally, data points could be distributed according to certain distribution $\{p^i\}_{i=1}^N$, and the mean is defined accordingly as:
\begin{align}
    \mu = \sum_{i=1}^N p^i \xbf^i
\end{align}
The centered dataset is formed by taking defining new coordinates for all data points:
\begin{align}
    b^i = \xbf^i - \mu
\end{align}
The covariance matrix $A$ is defined as:
\begin{align}
    A_{jk} = \sum_{i=1}^N p^i b_j^i (b_k^i)^* = \sum_{i=1}^N p^i \xbf^i_j (\xbf_k^i)^*-\mu_j\mu_k 
\end{align}
As been shown in \cite{gordon2022covariance}, the above quantity is exactly: 
\begin{align}
   \sum_{i=1}^N p^i b_j^i (b_k^i)^* =  \sum_{i=1}^N p^i \xbf^i_j (\xbf_k^i)^*-\mu_j\mu_k 
\end{align}
and hence $A = \sum_{i=1}^N p^i \xbf^i (\xbf^i)^\dagger - \mu \mu^\dagger $ where $(*)$ refers to the complex conjugate. The critical step of PCA is the diagonalization of matrix $A$, and the top eigenvalues/eigenvectors of $A$ are regarded as principal component analysis. For the given dataset $\{\xbf^i\}_{i=1}^N$, if instead of diagonalizing $A$ as defined above, we diagonalize the matrix defined by:
\begin{align}
    B_{ij} = \sum_{i=1}^N p^i \xbf^i_j (\xbf_k^i)^* 
\end{align}
which basically implies that $B = \sum_{i=1}^N p^i \xbf^i (\xbf^i)^T $. Then the method is called ``Principal component analysis without centering''. Detailed discussion and related rigorous construction of this PCA without centering  have been outlined in \cite{gordon2022covariance}. It turns out that PCA without centering is closely related to regular PCA and that the top eigenvalues/eigenvectors could effectively be used. Thus, it releases the requirement for centered data set, which might potentially makes it easier to implement in practice. However, the aim of us is to treat PCA in full generality, which includes the contribution from the centroids. 

A central step of PCA (both with and without centering) is the preparation of the so-called covariance matrix $A$ (or $B$). In the original work \cite{lloyd2013quantum}, the authors suggested a means of preparing such a matrix using quantum random access memory \cite{giovannetti2008architectures,giovannetti2008quantum}, as it gives the ability to coherently load classical data into the quantum state. However, a concrete procedure was lacking and the work of \cite{gordon2022covariance} have filled that gap. In their work, they assume that data points are given as quantum state, which is essentially a normalized vector $\{ \ket{\xbf^i}\}_{i=1}^N$ and for all $i$, $\ket{\xbf^i} \in \mathbb{C}^n$, prepared by amplitude encoding methods \cite{grover2000synthesis,grover2002creating, plesch2011quantum,nakaji2022approximate,marin2023quantum}, i.e., for each $\ket{\xbf^i}$ there exists a unitary $U_i$ having depth $\mathcal{O}(\log(n))$ such that $U_i \ket{\bf 0}  = \ket{\xbf^i}$. If, imagine a procedure whereby one sample from probability distribution $\{p^i\}_{i=1}^N$, and if the outcome is $i$ one uses the unitary $U_i$ to prepare the state $\ket{\xbf^i}$, then the result is an effective preparation of the mixed state 
$$ \Bar{\rho} = \sum_{i=1}^N p^i \ket{\xbf^i}\bra{\xbf^i}$$
which corresponds to the matrix $B$ defined above (more precisely, there is an extra normalization as assumed in \cite{gordon2022covariance}). Hence, PCA without centering could be executed simply with the above density state. 

In fact, given the prepared density state corresponds to the matrix $B$ above, one can execute our newly proposed quantum algorithm from section \ref{sec: quantumalgorithm} (and apparently, the original quantum PCA in \cite{lloyd2013quantum}) in a straightforward manner, to reveal the largest eigenvalues/eigenvectors of matrix $B$. However, as we mentioned, the scope of our work is to take into account centroids, e.g. the term $\mu \mu^T$ and to treat PCA in its full generality. Given (multiple copies of) the $\Bar{\rho}$ as above, one first needs to perform the simulation $\exp(-i \Bar{\rho} t)$ and then use the lemma \ref{lemma: logarithmicofunitary} to construct the block encoding of $\pi\Bar{\rho}/4$. In order to handle the centroids term $\mu \mu^T$, we need a means to prepare the block encoding of $\pi \mu \mu^T/4$. To achieve such a goal, remind that we are assumed to have unitaries $\{U_i\}_{i=1}^N$ that prepare the corresponding data points $\{ \ket{\xbf^i} \}_1^N$. Because $U_i \ket{0}^{\otimes \log(n)} = \ket{\xbf^i}$, then we have that $U_i$ contains $\ket{\xbf^i} \equiv \sum_{j=1}^n x^i_j \ket{j}$ as its first column, e.g.
\begin{align}
    U_i = \begin{pmatrix}
        x^i_1 & \cdot & \cdots & \cdot \\
        x^i_2 & \cdot & \cdots & \cdot \\
        \vdots & \cdot & \ddots & \cdot \\
        x^i_n & \cdot & \cdots & \cdot 
    \end{pmatrix}
\end{align}
As every unitary trivially block-encodes itself (see definition \ref{def: blockencode}), one can then use lemma \ref{lemma: sumencoding}, to construct the block encoding of $ \frac{1}{p}\sum_{i=1}^Np^iU_i$ (where $p = \sum_i p^i =1 $), using further $\mathcal{O}(\log(N))$ ancilla qubits, and a quantum circuit of depth $\mathcal{O}(N)$ (where we have assumed that each $U_i$ has depth $\mathcal{O}(1)$, i.e., independent of $N$). Given that $U_i$ has the form defined above, we see that the first column of $\sum_{i=1}^N p^i U_i$ is indeed $ \sum_{i=1}^N p^i \ket{\xbf^i} $. Let $U_p$ denotes the unitary block encoding of $\sum_{i=1}^N p^i U_i$. Then according to definition \ref{def: blockencode} (and also property below that) we have that:
\begin{align}
    U_p\ket{\bf 0}\ket{0}^{\otimes \log(M)} &= \ket{\bf 0} \big(  \sum_{i=1}^N p^i U_i \ket{0}^{\otimes \log(M)}  \big)  + \ket{\rm Garbage} \\
    &= \ket{\bf 0} \big( \sum_{i=1}^N p^i \ket{\xbf^i} \big)  + \ket{\rm Garbage} 
\end{align}
where $\ket{\rm Garbage}$ is some redundant state that is completely orthogonal to the first part $ \ket{\bf 0} \big( \sum_{i=1}^N p^i U_i \ket{0}^{\otimes \log(M)}  \big) $. 

Let $\ket{\Phi} \equiv \ket{\bf 0} \big( \sum_{i=1}^N p^i \ket{\xbf^i} \big)$. We have the following recipe derived from QSVT \cite{gilyen2019quantum}:
\begin{lemma}[\cite{gilyen2019quantum} Block Encoding of a Density Matrix]
\label{lemma: improveddme}
Let $\sigma = \Tr_A \ket{\Phi}\bra{\Phi}$, where $\sigma \in \mathbb{H}_B$, $\ket{\Phi} \in  \mathbb{H}_A \otimes \mathbb{H}_B$. Given unitary $U$ that generates $\ket{\Phi}$ from $\ket{\bf 0}_A \otimes \ket{\bf 0}_B$, then there exists a highly efficient procedure that constructs an exact unitary block encoding of $\rho$ using $U$ and $U^\dagger$ a single time, respectively.
\end{lemma}
Given that we have shown how to construct $U_p$, and that $U_p$ generates $\ket{\Phi}$, it is straightforward to apply the above lemma to construct the block encoding of $\ket{\Phi}\bra{\Phi}$. It is simple to see that the density matrix representation:
\begin{align}
    \ket{\Phi}\bra{\Phi} &= \big( \ket{\bf 0} \big( \sum_{i=1}^N p^i \ket{\xbf^i} \big)  + \ket{\rm Garbage}  \big)  \big(  \bra{\bf 0} \sum_{i=1}^N p^i \bra{\xbf^i} + \bra{\rm Garbage}   \big) \\
    &= \ket{\bf 0}\bra{\bf 0} \otimes  \mu \mu^\dagger + (...) 
\end{align}
where $(...)$ simply denotes the redundant terms, as we only pay attention to the first term. In fact, according to definition \ref{def: blockencode}, the above term is actually the block encoding of $\mu\mu^\dagger$, which means that using the lemma \ref{lemma: improveddme} to construct the block encoding of $\ket{\Phi}\bra{\Phi}$, we naturally obtain the block encoding of $\mu\mu^\dagger
$. Then lemma \ref{lemma: scale} could be employed to construct the block encoding of $\pi\mu\mu^\dagger/4$. Thus, given that we have already showed how to obtain the block encoding of $\pi \Bar{\rho}/4$, we can use lemma \ref{lemma: sumencoding} to construct the block encoding of
$$ \frac{\pi}{8} ( \Bar{\rho} - \mu\mu^\dagger)$$
With such a recipe, we can proceed with the algorithm developed in section \ref{sec: quantumalgorithm} to reveal the principal components of $\frac{\pi}{4} ( \Bar{\rho} - \mu\mu^\dagger) $, which is essentially the same as principal components of $\Bar{\rho} - \mu\mu^\dagger$ differed by a factor of $\pi/8$. 

We remark on one subtlety. As a first step, we prepare the mix state $\Bar{\rho}$ by sampling from distribution $\{p^i\}_{i=1}^N$ and accordingly prepare the state $\ket{\xbf^i}$ , resulting in an ensemble $\Bar{\rho} = \sum_{i=1}^N p^i \ket{\xbf^i}\bra{\xbf^i}$. Then we use lemma \ref{lemma: dme} and lemma \ref{lemma: unitaryblockencodingrho}  to obtain an $\epsilon$-approximated block encoding of $\pi\Bar{\rho}/4$ with complexity $\mathcal{O}(\log(n)/\epsilon^2)$. Then the step of preparing the (exact) block encoding of $\pi\mu\mu^\dagger/4$ takes a greater complexity $\mathcal{O}( N  )$, resulting in a total complexity $\mathcal{O}( \log(n)\frac{1}{\epsilon^2}+ N )$. It turns out that there is a way to trade the complexity between $1/\epsilon^2$ and $N$. Recall that for all $i$, we assume that we have unitaries $U_i$ (of depth $\mathcal{O}(\log(n))$) such that $U_i \ket{0}^{\log(n)} = \ket{\xbf^i}$. Then we can directly use the lemma \ref{lemma: improveddme} to construct an exact block encoding of $\ket{\xbf^i}\bra{\xbf^i}$, for all $i$. Then we can use lemma \ref{lemma: sumencoding} to construct an exact block encoding of the desired linear combination $\sum_{i=1}^N p^i \ket{\xbf^i}\bra{\xbf^i}$. The quantum circuit depth of this step is $\mathcal{O}(N \log(n))$. Thus, we have achieved an alternative route to prepare an exact block encoding of the desired mixed state $\Bar{\rho} = \sum_{i=1}^N p^i \ket{\xbf^i}\bra{\xbf^i}$, without error induced. We summarize the result of this section in the following lemma.
\begin{lemma}[Covariance Matrix Preparation]
    \label{lemma: covariancematrixpreparation}
    Let the dataset be $\{ \ket{\xbf^i} \}_{i=1}^N$ where each $\ket{\xbf^i} \in \mathbb{C}^n$. Define the centroid and ensembles as:
    \begin{align*}
        \mu = \sum_{i=1}^N p^i \ket{\xbf^i},  \Bar{\rho} = \sum_{i=1}^N p^i \ket{\xbf^i}\bra{\xbf^i}
    \end{align*}
    where $\{p^i\}^N_{i=1}$ is the probability distribution of a given dataset. There exist two procedures that achieve the following: \begin{itemize}
        \item A quantum circuit of depth $\mathcal{O}( N + \log(n)\frac{1}{\epsilon^2} )$ yielding an $\epsilon$-approximated block encoding of $\frac{\pi}{8}(  \Bar{\rho} - \mu \mu^\dagger )$
        \item A quantum circuit of depth $\mathcal{O}( N +  N  \log(n)  )$ yielding an exact block encoding of $\frac{\pi}{8}(  \Bar{\rho} - \mu \mu^\dagger )$
    \end{itemize}
\end{lemma}

\section{Outlook and Conclusion}
\label{sec: outlook}
In this work, we have used elegant techniques from the seminal quantum singular value transformation framework to shed new light on the problem of (quantum) principal component analysis. The central insight of our newly crafted algorithm relies on the definition of block encoding operators and that we can perform transformation on such a block-encoded operator, which is also the underlying power of QSVT. The exponentiation of a given density matrix turns out to be suitable to apply QSVT, as it is already a unitary matrix and by using a simple polynomial transformation, one can recover the block encoding of the density matrix itself, which further enables us to incorporate the quantum power algorithm proposed in \cite{nghiem2023improved,nghiem2023improved1}. As a result, the largest eigenvalue of a given density matrix could be revealed with the corresponding eigenvector. Extension to a few largest eigenvalues/eigenvectors turned out to be straightforward. Upon examining the complexity in detail, we came to the conclusion that our framework performs the best when the given density matrix $\rho$ exhibits considerable gaps between, for example, $R$ largest eigenvalues. At the same time, as discussed in \cite{lloyd2013quantum}, their quantum PCA algorithm achieves the best performance whenever there exists a top $R$ eigenvalues that are largely dominant from the remaining, and these eigenvalues are equal on average. 

Aside from performing quantum principal component analysis, we also revisit the problem of preparing the covariance matrix, which is a crucial part of PCA with respect to real-world application. As the problem has been touched on in \cite{gordon2022covariance}, our work uses QSVT to go deeper into the part where \cite{gordon2022covariance} did not fully justify, i.e., the contribution from centroids. Although the work \cite{gordon2022covariance} has provided a theoretical result quantifying how much ``PCA without centering'' differs from regular PCA, we can see that our framework can naturally deal with uncentered data, which makes it quite more useful for practical purposes. As we can see from the above lemma \ref{lemma: covariancematrixpreparation}, the second approach yields an error-free construction of the desired covariance matrix, which might be even more useful for subsequent application of quantum power method (as in Section \ref{sec: quantumalgorithm}, theorem \ref{theorem: improvedqpca}, \ref{theorem: principalcomponents}). Therefore, the second route is most efficient when $N$ is small, which means that we have a sufficiently small dataset. Overall, since the complexity of our quantum principal component analysis algorithm depends logarithmically on the dimension $n$ of the dataset, and linear to the number of data points $N$, we affirm that it would be most effective in the context of small pool, potentially very high dimensional data. 

Alongside the development of quantum machine learning algorithms, there exists a class of algorithms called quantum-inspired algorithms. In a series of efforts \cite{tang2019quantum,tang2021quantum, gilyen2018quantum}, the authors showed that under appropriate input assumption, there are efficient classical algorithms that could perform many tasks with at most a polynomial slowdown compared to the best quantum ones, thus resisting many previously claimed quantum speedups, including quantum principal component analysis. It is worth to mention that these quantum-inspired algorithms work best in the regime of low rank, meanwhile, we have pointed out that our method could still perform well in both low-rank and high-rank regime, as long as there is a sufficiently large enough gap between eigenvalues. Thus, we believe that the protocol introduced in this work, including both the new quantum principal component analysis algorithm as well as the new protocol for preparing the covariance matrix, is still gonna be effective for application outside of physical reign.

\section*{Acknowledgement}
The author acknowledges the support from the Center for Distributed Quantum Processing, Stony Brook University.

\bibliography{ref.bib}{}
\bibliographystyle{unsrt}

\clearpage
\newpage
\onecolumngrid
\appendix

\section*{Preliminaries}
\label{sec: prelim}
Here, we summarize the main recipes of our work, which mostly derived in the seminal QSVT work \cite{gilyen2019quantum}. We keep the statements brief and precise for simplicity, with their proofs/ constructions referred to in their original works.

\begin{definition}[Block Encoding Unitary]~\cite{low2017optimal, low2019hamiltonian, gilyen2019quantum}
\label{def: blockencode} 
Let $A$ be some Hermitian matrix of size $N \times N$ whose matrix norm $|A| < 1$. Let a unitary $U$ have the following form:
\begin{align*}
    U = \begin{pmatrix}
       A & \cdot \\
       \cdot & \cdot \\
    \end{pmatrix}.
\end{align*}
Then $U$ is said to be an exact block encoding of matrix $A$. Equivalently, we can write:
\begin{align*}
    U = \ket{ \bf{0}}\bra{ \bf{0}} \otimes A + \cdots
\end{align*}
where $\ket{\bf 0}$ refers to the ancilla system required for the block encoding purpose. In the case where the $U$ has the form 
$$ U  =  \ket{ \bf{0}}\bra{ \bf{0}} \otimes \Tilde{A} + \cdots $$
where $|| \Tilde{A} - A || \leq \epsilon$ (with $||.||$ being the matrix norm), then $U$ is said to be an $\epsilon$-approximated block encoding of $A$.
\end{definition}

The above definition has multiple natural corollaries. First, an arbitrary unitary $U$ block encodes itself. Suppose $A$ is block encoded by some matrix $U$. Next, then $A$ can be block encoded in a larger matrix by simply adding any ancilla (supposed to have dimension $m$), then note that $\Ibb_m \otimes U$ contains $A$ in the top-left corner, which is block encoding of $A$ again by definition. Further, it is almost trivial to block encode identity matrix of any dimension. For instance, we consider $\sigma_z \otimes \Ibb_m$ (for any $m$), which contains $\Ibb_m$ in the top-left corner. We further notice that from the above definition, the action of $U$ on some quantum state $\ket{\bf 0}\ket{\phi}$ is:
\begin{align}
    \label{eqn: action}
    U \ket{\bf 0}\ket{\phi} = \ket{\bf 0} A\ket{\phi} + \ket{\rm Garbage},
\end{align}
where $\ket{\rm Garbage }$ is a redundant state that is orthogonal to $\ket{\bf 0} A\ket{\phi}$.

\begin{lemma}[\cite{gilyen2019quantum} \revise{Block Encoding of a Density Matrix}]
\label{lemma: improveddme}
Let $\rho = \Tr_A \ket{\Phi}\bra{\Phi}$, where $\rho \in \mathbb{H}_B$, $\ket{\Phi} \in  \mathbb{H}_A \otimes \mathbb{H}_B$. Given unitary $U$ that generates $\ket{\Phi}$ from $\ket{\bf 0}_A \otimes \ket{\bf 0}_B$, then there exists a highly efficient procedure that constructs an exact unitary block encoding of $\rho$ using $U$ and $U^\dagger$ a single time, respectively.
\end{lemma}

The proof of the above lemma is given in \cite{gilyen2019quantum} (see their Lemma 45). \\



\begin{lemma}[Block Encoding of Product of Two Matrices]
\label{lemma: product}
    Given the unitary block encoding of two matrices $A_1$ and $A_2$, then there exists an efficient procedure that constructs a unitary block encoding of $A_1 A_2$ using each block encoding of $A_1,A_2$ one time. 
\end{lemma}

\begin{lemma}[\cite{camps2020approximate} \revise{Block Encoding of a Tensor Product}]
\label{lemma: tensorproduct}
    Given the unitary block encoding $\{U_i\}_{i=1}^m$ of multiple operators $\{M_i\}_{i=1}^m$ (assumed to be exact encoding), then, there is a procedure that produces the unitary block encoding operator of $\bigotimes_{i=1}^m M_i$, which requires \revise{parallel single uses} of 
    $\{U_i\}_{i=1}^m$ and $\mathcal{O}(1)$ SWAP gates. 
\end{lemma}
The above lemma is a result in \cite{camps2020approximate}. 
\begin{lemma}[\revise{\cite{gilyen2019quantum} Block Encoding of a  Matrix}]
\label{lemma: As}
    Given oracle access to $s$-sparse matrix $A$ of dimension $n\times n$, then an $\epsilon$-approximated unitary block encoding of $A/s$ can be prepared with gate/time complexity $\mathcal{O}\Big(\log n + \log^{2.5}(\frac{s^2}{\epsilon})\Big).$
\end{lemma}
This is presented in~\cite{gilyen2019quantum} (see their Lemma 48), and one can also find a review of the construction in~\cite{childs2017lecture}. We remark further that the scaling factor $s$ in the above lemma can be reduced by the preamplification method with further complexity $\mathcal{O}({s})$~\cite{gilyen2019quantum}.

\begin{lemma}[\cite{childs2012hamiltonian, gilyen2019quantum} Linear combination of block-encoded matrices]
    Given unitary block encoding $U_i$ of multiple operators $\{M_i\}_{i=1}^m$ (with complexity $T_M$) and the ability to implement controlled-$U_i$ for all $i$. Given further the ability to prepare a state $\frac{1}{\sqrt{\alpha}}\sum_{i=1}^{m} \sqrt{\alpha_i}\ket{i-1}$ where $\alpha = \sum_{i=1}^{m} \alpha_i$, then there is a quantum circuit of depth $\mathcal{O}(m T_M)$ that produces a unitary block encoding operator of \,$\sum_{i=1}^m \pm \alpha_i M_i/\alpha $. 
    \label{lemma: sumencoding}
\end{lemma}

\begin{lemma}[Scaling Block encoding] 
\label{lemma: scale}
    Given a block encoding of some matrix $A$ (as in~\ref{def: blockencode}), then the block encoding of $A/p$ where $p > 1$ can be prepared with an extra $\mathcal{O}(1)$ cost.  
\end{lemma}
To show this, we note that the matrix representation of RY rotational gate is
\begin{align}
   R_Y(\theta) = \begin{pmatrix}
        \cos(\theta/2) & -\sin(\theta/2) \\
        \sin(\theta/2) & \cos(\theta/2) 
    \end{pmatrix}.
\end{align}
If we choose $\theta$ such that $\cos(\theta/2) = 1/p$, then Lemma~\ref{lemma: tensorproduct} allows us to construct block encoding of $R_Y(\theta) \otimes \mathbb{I}_{{\rm dim}(A)}$  (${\rm dim}(A)$ refers to dimension of matirx $A$), which contains the diagonal matrix of size ${\rm dim}(A) \times {\rm dim}(A)$ with entries $1/p$. Then Lemma~\ref{lemma: product} can construct block encoding of $(1/p) \ \mathbb{I}_{{\rm dim}(A)} \cdot A = A/p$.  \\

The following is called amplification technique:
\begin{lemma}[\cite{gilyen2019quantum} Theorem 30; \revise{\bf Amplification}]\label{lemma: amp_amp}
Let $U$, $\Pi$, $\widetilde{\Pi} \in {\rm End}(\mathcal{H}_U)$ be linear operators on $\mathcal{H}_U$ such that $U$ is a unitary, and $\Pi$, $\widetilde{\Pi}$ are orthogonal projectors. 
Let $\gamma>1$ and $\delta,\epsilon \in (0,\frac{1}{2})$. 
Suppose that $\widetilde{\Pi}U\Pi=W \Sigma V^\dagger=\sum_{i}\varsigma_i\ket{w_i}\bra{v_i}$ is a singular value decomposition. 
Then there is an $m= \mathcal{O} \Big(\frac{\gamma}{\delta}
\log \left(\frac{\gamma}{\epsilon} \right)\Big)$ and an efficiently computable $\Phi\in\mathbb{R}^m$ such that
\begin{equation}
\left(\bra{+}\otimes\widetilde{\Pi}_{\leq\frac{1-\delta}{\gamma}}\right)U_\Phi \left(\ket{+}\otimes\Pi_{\leq\frac{1-\delta}{\gamma}}\right)=\sum_{i\colon\varsigma_i\leq \frac{1-\delta}{\gamma} }\tilde{\varsigma}_i\ket{w_i}\bra{v_i} , \text{ where } \Big|\!\Big|\frac{\tilde{\varsigma}_i}{\gamma\varsigma_i}-1 \Big|\!\Big|\leq \epsilon.
\end{equation}
Moreover, $U_\Phi$ can be implemented using a single ancilla qubit with $m$ uses of $U$ and $U^\dagger$, $m$ uses of C$_\Pi$NOT and $m$ uses of C$_{\widetilde{\Pi}}$NOT gates and $m$ single qubit gates.
Here,
\begin{itemize}
\item C$_\Pi$NOT$:=X \otimes \Pi + I \otimes (I - \Pi)$ and a similar definition for C$_{\widetilde{\Pi}}$NOT; see Definition 2 in \cite{gilyen2019quantum},
\item $U_\Phi$: alternating phase modulation sequence; see Definition 15 in \cite{gilyen2019quantum},
\item $\Pi_{\leq \delta}$, $\widetilde{\Pi}_{\leq \delta}$: singular value threshold projectors; see Definition 24 in \cite{gilyen2019quantum}.
\end{itemize}
\end{lemma}
\end{document}